\documentclass[11pt]{article}
\usepackage{latexsym,amssymb,enumerate,amsmath}

\newtheorem{theorem}{Theorem}[section]

\newcommand{\bel}{\begin{equation}\label}
\newcommand{\ee}{\end{equation}}
\newcommand{\beq}{\begin{eqnarray}\label} 
\newcommand{\eeq}{\end{eqnarray}}
\newcommand{\bc}{\begin{center}}
\newcommand{\ec}{\end{center}}
\newcommand{\non}{\nonumber}
\newcommand{\bom}{\mbox{\boldmath$\omega$}}
\newcommand{\I}{\int_{\mathcal{V}}}

\newcommand{\bu}{\mbox{\boldmath$u$}}
\newcommand{\bhu}{\mbox{\boldmath$\hat{u}$}}
\newcommand{\bx}{\mbox{\boldmath$x$}}
\newcommand{\br}{\mbox{\boldmath$r$}}
\newcommand{\bdf}{\mbox{\boldmath$f$}}
\newcommand\shalf{\ensuremath{{\scriptstyle\frac{1}{2}}}}

\begin{document}
\bc
\textbf{A hierarchy of length scales for weak solutions of the three-dimensional 
Navier-Stokes equations}
\ec  
\bc
\textbf{J. D. Gibbon}\footnote{Department of Mathematics, Imperial College London, 
SW7 2AZ, UK. (j.d.gibbon@ic.ac.uk).}
\ec

\pagestyle{myheadings} \markright{Navier-Stokes length scales (28th May 2011)}

\begin{abstract}
Moments of the vorticity are used to define and estimate a hierarchy of time-averaged inverse 
length scales for weak solutions of the three-dimensional, incompressible Navier-Stokes equations 
on a periodic box. The estimate for the smallest of these inverse scales coincides with the inverse 
Kolmogorov length but thereafter the exponents of the Reynolds number rise rapidly for correspondingly 
higher moments. The implications of these results for the computational resolution of small scale 
vortical structures are discussed.
\end{abstract}

\par\noindent
\textbf{Keywords\,:} {\small Navier-Stokes, weak solutions, moments of vorticity, inverse length scales}\\

\bc
\textit{Dedicated to David Levermore on the occasion of his 60th birthday.\\
To appear in Comm. Math. Sci.}
\ec

\section{\large Introduction}\label{sect1}

Resolution issues in computations of solutions of the three-dimensional Navier-Stokes 
equations are not only closely associated with the problem of regularity but they also 
raise the question of how resolution length scales can be defined and estimated. The 
Kolmogorov school of statistical turbulence suggests that a system of volume $L^3$ has 
a cut-off in its $-5/3$ energy spectrum at $k_{c} = \lambda_{k}^{-1} \sim L^{-1}Re^{3/4}$, 
which is known as the inverse Kolmogorov length. The wave-numbers $k > k_{c}$ are 
considered to lie in what is called the dissipation range \cite{Frischbk,Davidson04,Tsinober09}. 
Significant energy lying in this range can provoke intermittent events in the vorticity and 
strain fields characterized by violent, spiky departures away from space-time averages whose 
corresponding statistics appear to be non-Gaussian in character 
\cite{BT49,Kuo71,MS91,Sreenirev,Lathrop03}, although intermittent events may also be 
associated with the inertial range ($k< k_{c}$). Whether significant energy actually 
cascades down to the micro/nano-scales where the equation fails to be a valid model is 
intimately entwined not only with the open question of regularity but also with the role 
of the Navier-Stokes equations as a limit of kinetic theory \cite{BGL1,BGL2}. This phenomenon 
continues to pose severe computational challenges \cite{FO90,KO93,VM94,SED10}.  In 
statistical physics the objects that are used to study intermittency are the 
ensemble-averaged velocity structure functions
\bel{intermit}
\left<\,\left|u(\bx + \br) - u(\bx)\right|^{p}\,\right>_{en\!s.av.}\sim r^{\zeta_{p}}
\ee 
the departure of whose exponents $\zeta_{p}$ from linear\footnote{Kolmogorov predicted a 
linear relation between $\zeta_{p}$ and $p$\,: departure from this is called `anomalous 
scaling' and is usually manifest by $\zeta_{p}$ lying on a concave curve below linear 
for $p > 3$ \cite{Frischbk}. The two coincide for $p=3$.} is thought to be caused by 
inertial range intermittent behaviour \cite{Frischbk,Davidson04,Tsinober09}. It 
is clear, however, that these structure functions are not best suited for Navier-Stokes 
analysis\,: the task of this paper is to discuss what could replace these in the Navier-Stokes 
context and what information could be gleaned from them. While higher gradients of the velocity 
field would undoubtedly capture intermittent behaviour, 
they would be unreachable computationally for all practical purposes. A better diagnostic of 
spikiness in Navier-Stokes solutions is a sequence of $L^p$-norms, or higher moments, of the 
vorticity $\bom = \mbox{curl}\,\bu$, defined through the set of frequencies ($p=2m$ for $m > 1$) 
\bel{Omdef}
\Omega_{m}(t) = \left(L^{-3}\I |\bom|^{2m}dV\right)^{1/2m}\,,
\ee
where the domain $\mathcal{V} = [0,\,L]^{3}$ is taken to be periodic. The basic frequency 
associated with the domain is given by $\varpi_{0} = \nu L^{-2}$\,. $\Omega_{1}^2$ is the 
enstrophy per unit volume which is related to the energy dissipation rate, whereas the higher 
moments will naturally pick up events at smaller scales. 
\par\smallskip
The setting is the incompressible ($\rm{div}\,\bu = 0$), forced, three-dimensional Navier-Stokes 
equations for the velocity field $\bu(\bx,\,t)$
\bel{NS1}
\partial_{t}\bu + \bu\cdot\nabla\bu = \nu\Delta\bu - \nabla p + \bdf(\bx)\,.
\ee
Traditionally, most estimates in Navier-Stokes analysis have been found in terms of the Grashof number 
$Gr$, which is expressed in terms of the root mean square ($f_{rms}^{2} = L^{-3}\|\bdf\|_{2}^{2}$) of 
the divergence-free forcing $\bdf(\bx)$ (see \cite{CF,Temam,FMRT,DGbook}) but it would be more helpful 
to express these in terms of the Reynolds number $Re$ to facilitate comparison with the results of 
statistical physics. The definitions of $Gr$ and $Re$ are
\bel{GrRedef}
Gr = L^{3}f_{rms}\nu^{-2}\,,\qquad\qquad Re = U_{0}L \nu^{-1}\,.
\ee
Doering and Foias \cite{DF02} used the idea of defining $U_{0}$ as
\bel{U0def}
U_{0}^{2} = L^{-3}\big<\|\bu\|_{2}^{2}\big>_{T}
\ee
where the time average $\big<\cdot\big>_{T}$ over an interval $[0,\,T]$ is defined by
\bel{timeavdef}
\left<g(\cdot)\right>_{T} = \limsup_{g(0)}\frac{1}{T}\int_{0}^{T}g(\tau)\,d\tau\,.
\ee
Clearly, $Gr$ is fixed provided $\bdf$ is $L^2$-bounded, while $Re$ is the system response 
to this forcing. A brief look at Leray's energy inequality shows why this definition of $U_{0}$ 
is of value
\bel{Ler1}
\shalf \frac{d~}{dt}\I |\bu|^{2}\,dV \leq -\nu \I |\bom|^{2}\,dV + \|\bdf\|_{2}\|\bu\|_{2}\,,
\ee
because it leads to
\bel{Ler3}
\left<\Omega_{1}^{2}\right>_{T} \leq \varpi_{0}^{2}Gr\,Re + O\left(T^{-1}\right)\,.
\ee
With some very mild technical restrictions on the forcing\footnote{Doering and Foias \cite{DF02} 
took narrow-band forcing around a specific wave-number but a wave-number spectrum which is cut off 
both above and below is sufficient.}, Doering and Foias \cite{DF02} then showed that Navier-Stokes 
solutions obey $Gr \leq c\,Re^{2}$. This turns (\ref{Ler3}) into
\bel{Ler4}
\left<\Omega_{1}^{2}\right>_{T} \leq c\,\varpi_{0}^{2}Re^{3} + O\left(T^{-1}\right)\,.
\ee
In fact $\nu\left<\Omega_{1}^{2}\right>_{T}$ is the time-averaged energy dissipation rate 
per unit volume over $[0,\,T]$ and allows us to form and bound from above the inverse 
Kolmogorov length scale $\lambda_{k}^{-1}$
\bel{Koldef1}
\lambda_{k}^{-4} = \frac{\nu\left<\Omega_{1}^{2}\right>_{T}}{\nu^{3}}
\qquad\Rightarrow\qquad L\lambda_{k}^{-1}\leq c\,Re^{3/4} + O\big(T^{-{1/4}}\big)\,.
\ee
An estimate for the inverse Taylor micro-scale $\lambda_{Tms}^{-1}$ can also be found from (\ref{Ler4})
\bel{tms1}
L\lambda_{Tms}^{-1} := 
L\left(\frac{\left<\|\bom\|_{2}^{2}\right>_{T}}{\left<\|\bu\|_{2}^{2}\right>_{T}}\right)^{1/2}
\leq c\,Re^{1/2}  + O\big(T^{-{1/2}}\big)\,.
\ee  
Both these upper bounds gratifyingly coincide with the results of statistical turbulence 
theory \cite{Frischbk,Davidson04,Tsinober09} although the fact that they are bounds allows 
for structures to occur in a flow whose natural scales are larger \cite{VH91}.  The question 
is now clear\,: \textit{can we construct and bound from above a sequence of inverse length 
scales associated with the higher moments $\Omega_{m}$?}

\section{\large A scaling property and length scale estimates}\label{sect2}

Leray's energy inequality (\ref{Ler1}) is valid for weak solutions and thus the estimate (\ref{Ler4}) 
is rigorous, although the existence and uniqueness of solutions for arbitrarily long times remain an 
open problem. While it is 
possible to subscribe to the view that difficulties in flow resolution could be a symptom of the lack 
of uniqueness of weak solutions, in tandem it ought also to be acknowledged that these difficulties 
may simply be caused by the practical challenges of working on a system where even the naturally 
largest scale (other than $L$) lies close to the limit of what can currently be resolved. The spirit 
of this paper is such that results on weak solutions are assumed to be sufficiently physical to reflect 
the reality of turbulent flows, provided $T$ is taken large enough\footnote{While existence and uniqueness 
of solutions is easily proved for small values of $T$ \cite{CF,Temam,FMRT}, larger values than this are 
necessary to make sense of long-time averages.}. This strategy allows the estimation of an infinite 
hierarchy of time averages of powers of the $\Omega_{m}$ for weak solutions on $[0,\,T]$ without 
having to appeal to point-wise estimates that the solution of the regularity problem would require. 
In turn, these time averages allow us to define and explore the natural length scales inherent in 
the system. The following result was proved in \cite{JDGPRS10} under the assumption 
that strong solutions exist. Here it is demonstrated for weak solutions\,:
\par\vspace{1mm}
\begin{theorem}\label{thm1}
Weak solutions of the three dimensional Navier-Stokes equations satisfy 
\bel{FGT1}
\left<\left(\varpi_{0}^{-1}\Omega_{m}\right)^{\alpha_{m}}\right>_{T} \leq c\,Re^{3} + O\big(T^{-1}\big)\,,
\qquad\qquad 1 \leq m \leq \infty\,,
\ee
where $c$ is a uniform constant and 
\bel{alphamdef}
\alpha_{m} = \frac{2m}{4m-3}\,.
\ee
\end{theorem}
\textbf{Remark\,:} The exponent $\alpha_{m} = 2m/(4m-3)$ within (\ref{FGT1}) appears as a natural 
scaling of the Navier-Stokes equations, consistent with the application of H\"{o}lder and Sobolev 
inequalities. Note that when $m=1$ the value $\alpha_{1} = 2$ is consistent with (\ref{Ler4}). 
\par\medskip\noindent
\textbf{Proof\,:} The proof is based on a result of Foias, Guillop\'e and Temam \cite{FGT} 
(Theorem 3.1) for weak solutions which, when modified in the manner of Doering and 
Foias \cite{DF02}, furnishes us with the following time averaged estimate 
\bel{FGT1a}
\left< H_{N}^{\frac{1}{2N-1}}\right>_{T} \leq c_{N}L^{-1}\nu^{\frac{2}{2N-1}}Re^{3} + O\left(T^{-1}\right)\,,
\ee
where
\bel{HNdef}
H_{N} = \I \left|\nabla^{N}\bu\right|^{2}\,dV = \int_{\mathcal{V}_{k}} k^{2N}\left|\bhu\right|^{2}\,d^{3}k\,,
\ee
where $H_{1} = \I \left|\nabla\bu\right|^{2}\,dV = \I \left|\bom\right|^{2}\,dV$. An interpolation 
between $\|\bom\|_{2m}$ and $\|\bom\|_{2}$ is found using $H_{N}$ 
\bel{FGT2} 
\|\bom\|_{2m} \leq c_{N,m} \|\nabla^{N-1}\bom\|_{2}^{a}\,\|\bom\|_{2}^{1-a}\,,\qquad\qquad  
a = \frac{3(m-1)}{2m(N-1)}\,,
\ee
for $N\geq 3$. $\|\bom\|_{2m}$ is now raised to the power $\alpha_{m}$, which is to be determined, 
and the time average of this is estimated as
\beq{FGT3}
\left<\|\bom\|_{2m}^{\alpha_{m}}\right>_{T} &\leq& 
c_{N,m}^{\alpha_{m}} \left<\|\nabla^{N-1}\bom\|_{2}^{a\alpha_{m}}
\|\bom\|_{2}^{(1-a)\alpha_{m}}\right>_{T}\non\\
&=& c_{N,m}^{\alpha_{m}} \left<\left(H_{N}^{\frac{1}{2N-1}}\right)^{\shalf a\alpha_{m}(2N-1)}
H_{1}^{\shalf(1-a)\alpha_{m}}\right>_{T}\non\\
&\leq& c_{N,m}^{\alpha_{m}} \left<H_{N}^{\frac{1}{2N-1}}\right>_{T}^{\shalf a\alpha_{m}(2N-1)}
\left<H_{1}^{\frac{(1-a)\alpha_{m}}{2-a\alpha_{m}(2N-1)}}\right>_{T}^{1-\shalf a\alpha_{m}(2N-1)}
\eeq
An explicit upper bound in terms of $Re$ is available only if the exponent of $H_{1}$ within the 
average is unity\,; that is
\bel{FGT4}
\frac{(1-a)\alpha_{m}}{2-a\alpha_{m}(2N-1)} = 1\,.
\ee
This determines $\alpha_{m}$, uniformly in $N$, as
\bel{FGT5}
\alpha_{m} = \frac{2m}{4m-3}\,.
\ee
Using the estimate in (\ref{FGT1a}), and that for $\left<H_{1}\right>$, the result follows. 
The constant $c_{N,m}$ can be minimized by choosing $N=3$. $c_{3,m}$ does not blow up even 
when $m=\infty$\,; thus we take the largest value of $c_{3,m}^{\alpha_{m}}$ and call this 
$c$. \hfill $\blacksquare$

\section{\large Definition of the inverse length scales}

Based on the definition of the inverse Kolmogorov length $\lambda_{k}^{-1}$ in (\ref{Koldef1}) 
a generalization of this to a hierarchy of inverse lengths $\lambda_{m}^{-1}$ suggests the definition 
\bel{lamdef1}
\big(L\lambda_{m}^{-1}\big)^{2\alpha_{m}} := \left<\left(\varpi_{0}^{-1}\Omega_{m}\right)^{\alpha_{m}}\right>_{T}\,.
\ee
The $\lambda_{m}$ are interpreted as resolution lengths in the space-time averaged sense for $1\leq m \leq \infty$\,:
\bel{lamdef2}
L\lambda_{m}^{-1} \leq c\,Re^{3/2\alpha_{m}} + O 
\left(T^{-1/2\alpha_{m}}\right)\,.
\ee
\par\vspace{-2mm}\noindent
Many turbulent structures have natural inverse gross length scales lying in the range between $Re^{1/2}$ 
and $Re^{3/4}$, but crinkles forming at finer scales may ultimately grow to be dominant and then become 
the cause of resolution difficulties \cite{Frischbk,Davidson04,Tsinober09,VM94,SED10}. For $m>1$ 
the $\lambda_{m}$ are interpreted here as the length scales corresponding to these deeper 
intermittent events. The upper bounds displayed in (\ref{lamdef2}), as the Table shows, range from 
the inverse Kolmogorov length $Re^{3/4}$ at $m=1$ to $Re^{3}$ for $m=\infty$. Computationally it 
may be hard to get far beyond $m=1$\,: for example, $m = 9/8$ corresponds to $Re^{1}$, which is close 
to modern resolutions even for modest values of $Re$. Thereafter the rise in the exponent $3/2\alpha_{m}$ 
is steep. Indeed, in the very high $m$ limit, the $Re^3$ bound has an exponent four times greater than 
the Kolmogorov length\,; this lies well below molecular scales where the equations are invalid. 
\par\vspace{0mm}\noindent
\begin{table}[htb]
\bc
\begin{tabular}{||c|c|c|c|c|c|c|c||}\hline
$m$ & 1 & 9/8 & 3/2 & 2 & 3 & \ldots & $\infty$\\\hline
$3/2\alpha_{m}$ & 3/4 & 1 &  3/2 & 15/8 & 9/4 & \ldots & 3\\\hline
$d_{m}$         &  3  & 2 &   1  &  3/5   & 1/2   & \ldots & 0 \\\hline
\end{tabular}
\caption{\textit{\small Values of the $Re$-exponent 
$3/2\alpha_{m} = 3\left(1 - \frac{3}{4m}\right)$, and $d_{m} = \frac{3}{4m-3}$.}}
\label{table1}
\ec
\end{table}
\par\vspace{0mm}\noindent
An interesting question is how the existence of this continuum of finer scales might 
be interpreted physically? To do so rigorously without a regularity proof is difficult 
but a very informal physical interpretation is possible in terms of the familiar concept 
of a cascade. One of the simplest cascade models is the so-called $\beta$-model of 
Frisch, Sulem and Nelkin \cite{FSN78} who, in analogy with Mandelbrot's ideas 
\cite{Mandel74}, modelled a Richardson cascade by taking a vortex of scale $\ell_{0} 
\equiv L$ and then allowed a cascade of daughter vortices, each of scale $\ell_{n}$. 
The idea was based on domain halving at each step such that $\ell_{0}/\ell_{n} = 2^{n}$. 
The self-similarity dimension $d$ 
(similar to fractal dimension) was then introduced by considering the number of offspring 
at each step as $2^{d}$\,: 2 for the halving of a line\,; 4 for the halving of each 
direction in the plane\,; and likewise 8 for the cube.  $d$ is then formally allowed 
to take non-integer values. In $d$ dimensions the corrections to the usual Kolmogorov 
scaling calculations for velocity, turn-over time etc appeared as multiplicative factors 
proportional to $\left(\ell_{0}/\ell_{n}\right)^{(3-d)/3}$\,: see \cite{Frischbk,FSN78}.  
Equating the turn-over and viscous times in the standard manner one arrives at ($\ell_{d}$ 
is their viscous dissipation length)\footnote{The integer $n$ labels the cascade\,: 
$m$ labels the higher moments as in (\ref{Omdef}), but does not necessarily take integer 
values. One choice is to take $n=m$ which means that each moment would correspond to a 
step in the cascade.  In \cite{FSN78} $d$ is not specified but is illustrative of 
the calculation necessary when applying Kolmogorov's theory in non-integer dimensions.}
\bel{Kold1}
\ell_{0}/\ell_{d} \sim Re^{\frac{3}{d+1}}\,.
\ee
This gives the usual Kolmogorov inverse scale of $Re^{3/4}$ in a fully three-dimensional 
domain but is shifted upwards for smaller values of $d$. Taking this idea as our physical 
analogy we compare (\ref{Kold1}) to the upper bound in (\ref{lamdef2}) to get 
\bel{Kold2}
d_{m}+1  = 2\alpha_{m}\qquad\Rightarrow\qquad d_{m} = \frac{3}{4m-3}\,,
\ee
where an $m$-label has been appended to $d$. Thus we are able to assign a corresponding 
self-similarity dimension $d_{m}$ to lower-dimensional vortical structures that require 
values of $m > 1$ for their resolution. Note that $d_{m}$ never goes negative. Models 
more sophisticated than the $\beta$-model, such as the bi-fractal and multi-fractal 
models \cite{Frischbk}, are more difficult to use as analogies as they would require 
data fitting.

\end{document}